# Nonequilibrium Critical Phenomena

Alexander Patashinski
Northwestern University, Evanston, IL 60208

*ABSTRACT*

We discuss the non-equilibrium critical phenomena in liquids, and the models for these phenomena based on local equilibrium and extended scaling assumptions. Special situations are proposed for experimental tests of the theory. Near-critical steady and transient states are reviewed. In a near-critical steady state characterized by a temperature gradient, the theory predicts strong nonequilibrium fluctuations at very large length scales. Close to the critical point, this results in a nonlinear regime of heat conductivity. A transient non-equilibrium state triggered by a rapid and large spatially uniform perturbation of the critical liquid is considered. A step away from criticality generates a free field with strong and decaying correlations in initial state, while a step towards criticality initiates the increase of fluctuations and of their correlation at the large scale edge of the critical range. The approach to equilibrium is characterized by an equilibration length $\lambda_{eq}$ that depends on time t. The theory predicts a power law approach of the temperature to the new equilibrium; the new critical exponents depend on whether the temperature is initially increased or decreased.

## *1. Motivation and objectives of the study*

The aim of this article is to discuss nonequilibrium critical phenomena in liquids and theoretical models for these phenomena, and to review some of recently obtained theoretical results. The critical point is the most studied example of field systems with strong interaction. These systems also include hydrodynamic turbulence, quantum vacuum (the arena of elementary particle physics), and other important systems; at present, our knowledge of these systems is insufficient even to classify them as belonging to same or different universality classes. The scaling theory and the renormalization group (RG) method [1-3] explain the observed thermodynamic and kinetic singularities in critical systems differing in the nature and symmetry properties of the fluctuating order parameter field. The foundations of the theory are, however, not rigorously derived from first principles. The scaling hypothesis is an assumption leading to a logically consistent closed theory well confirmed by the experiment.

The scaling theory suggests power lows for binary correlation functions, and relates the behavior of higher correlation functions to that of pair correlation function; a test of the latter part of the scaling hypothesis had not been done jet. Recent studies show that higher correlation functions in hydrodynamic turbulence may deviate from the predictions of a simple scaling theory [4] (anomalous scaling). Currently, there is little understanding of possible criteria for a scaling system to be normal or anomalous; in homogeneous equilibrium critical state, scaling is supposed to be normal. Ideally, one would like to continuously change some parameters of the system while keeping other constant, to arrive at a change of scaling behavior. A study of critical phenomena in non-uniform and non-equilibrium liquids is a step in this direction, providing for opportunities to get a deeper insight into mechanisms and limits of the general scaling behavior.

The uniform equilibrium critical state is uniquely defined by its temperature $T_c$ and pressure $P_c$. There are infinitely many ways to perturb this state to make it non-uniform and/or non-equilibrium. At the present, rather initial, stage of the study, one of the goals is to identify physical situations that may serve as standard for an experimental setup. In the following sections, we review physical situations suitable for experimental tests, and discuss theoretical models for non-equilibrium critical states. The consideration is limited to critical phenomena in liquid systems characterized by a scalar order parameter. The list of those systems includes liquid-gas critical points and binary solution (miscibility) critical points, and possibly the liquid-liquid phase transition criticality.

## *2. Near-critical states as perturbed critical state*

To describe nonequilibrium near-critical states, a generalization is needed of the models describing the equilibrium fluctuations. The critical state is characterized by strongly correlated fluctuations of the order parameter $\varphi(r,t)$ and other hydrodynamic fields [1-5] at length scales r ranging from atomic $r_0 \sim 10^{-7}$cm up to the size of the system which we consider infinite. According to the scaling hypothesis, at any large scale R the fluctuations are characterized by the same universal, non-Gaussian probability distribution function for the scaled order parameter field $\varphi/\varphi_R$, where $\varphi_R = Z(R)\varphi$. The dependence of the renormalization factor $Z(R)=R^{-(1+\eta)/2}$ defines the critical exponent $\eta$ [1-3]. Each length scale R is characterized by this scale relaxation time $\tau(R)$; the theory predicts the scaling low for the relaxation time as $\tau(R)=\tau(r_0)(R/r_0)^{(3\nu+z)/\nu}$, where z is the "dynamic" exponent. This critical point regime is rather fragile; a deviation of conditions from critical results, as a rule, in large changes of the fluctuation picture at some large length and time scales [1-3,5].

Near-critical states, including the nonequilibrium and non-uniform states, may be treated as resulting from a perturbation of the equilibrium uniform critical state. A special example is a uniform quasistatic deviation of the temperature and/or pressure from their critical values $T_c$, $P_c$. The perturbed state is then an equilibrium and uniform near critical state, characterized by a finite correlation radius $r_c(P-P_c, T-T_c)$ of fluctuations. On the critical isochore $V=V_c$, the correlation radius depends on temperature as $r_c(\varepsilon) = r_0 \varepsilon^{-\nu}$, where $\varepsilon = (T-T_c)/T_c$ is the reduced temperature, and $\nu$ the critical exponent for the correlation radius. At length scales smaller than $r_c$, the characteristics of fluctuations are, up to small corrections, the same as at the critical point, while at length scales larger than $r_c$ the interaction of fluctuations becomes weak and the statistics is Gaussian. This relatively higher stability of smaller scale fluctuations may be suggested as a general feature of thermal fluctuations, supporting the idea of a local equilibrium. The correlation times of equilibrium fluctuations in the critical range $r_0 < R < r_c$ of length scales range from microscopic times $\tau(r_0) \sim 10^{-12}$-$10^{-13}$ s up to the relaxation time of the correlation radius length scale $\tau(r_c) = \tau(r_0) \varepsilon^{-(3\nu+z)}$, where z is the "dynamic" critical exponent. The experimental benchmark, in microgravity conditions, is $\varepsilon \sim 10^{-8}$, $r_c \sim 10^{-3}$ cm, $\tau_c \sim 10$ sec [6]. The scaling formula for $\tau(R)$ reflects the strong interaction of equilibrium fluctuations in the critical range of scales $r_c > R > r_0$. At length scales $R > r_c$ this formula becomes inapplicable; fluctuations at this scales are Gaussian.

A more general type of perturbations leads to a nonequilibrium steady state. An example is a steady shear flow in a critical liquid; the flow disrupts the correlations at large length scales, resulting in non-isotropy of correlations and a diminishing the correlation radius [7,8]. To keep the state steady, the heat generated by the viscous energy loss in the bulk of the liquid has to be transferred to the boundary; the resulting inhomogeneity of the temperature is usually neglected in the theory.

Another non-equilibrium steady state appears when a temperature gradient is maintained in the liquid [9] by a difference in boundary temperatures. In this case, discussed in the next sections, a new range of nonequilibrium fluctuations appears at very large scales, and the heat exchange at the boundary results in a non-equilibrium thermal boundary layer[10]. The dominant mechanism of these effects is the mode coupling. In these cases, the perturbation of the critical point regime results in changes at large scales while the shorter scales are not changed significantly.

When the rate of changes of thermodynamic parameters (pressure P, temperature T) is finite, a transient state appears. A rapid change triggers relaxation to a new equilibrium state[11-12]. In this case of time inhomogeneity, the perturbation may be spatially uniform. The correlation radii of the initial (before change) and final state differ, and a range of critical correlations has to appear or disappear in the course of relaxation. In contrast to this large change at large scales, the fluctuation picture at small length scales $r<\min\{r_{c,init}, r_{c,fin}\}$ is only weakly perturbed by this rapid change of parameters. This different sensitivity of the fluctuation picture to perturbations is related to the local equilibrium, and is the basic assumption in the study of perturbation of the critical regime.

In the near-critical steady states caused by a heat and/or mass current, one deals with violations of space nonuniformity or non-isotropy, while the system is homogeneous in time. The transient states caused by a rapid uniform change of parameters are spatially uniform, but there is a time non-uniformity. In a more general case one deals with both space and time inhomogeneities acting simultaneously and interacting. This interaction can make the study and understanding of governing mechanisms very challenging. We propose to study the steady state with temperature gradient and the uniform transient states triggered by a rapid uniform perturbation as possible standard situations for further study.

At microscopic length scales R~$r_0$, the correlations depend on the microscopic details of interactions. At larger length scales, the fluctuation behavior is more universal, and may be described in terms of hydrodynamic fields. In the hydrodynamic description, details referring to length scales smaller than some *initial* cut-off scale $l_0 \gg r_0$ are averaged out. Once the theory is formulated in terms of hydrodynamic modes at some initial scale, one can use this theory to increase the cut-off length by eliminating (averaging out) the degrees of freedom referring to length scales between $l_0$ and $l>l_0$. One derives then new equations for length scales $l> l_{01}$. The length $l$ is now the new cut-off length. The changes, at increasing cut-off scale $l$, in the equations governing the hydrodynamic modes are considered in the RG approach.

Modeling of the perturbed critical state fluctuations and relaxation is made easier if the perturbation at the initial length scale may be regarded as small. Then, one may describe the perturbation as a small change in the governing equations and boundary conditions. Using the RG procedure, one calculates then the changes at larger length scales; these changes may appear large.

### *3. Langevin equations for non-equilibrium near-critical states*
The order parameter $\varphi(r,t)$ for the liquid-gas critical point is proportional to the deviation of the density and energy density from their critical values [2,3]; $\varphi(r)$ is a hydrodynamic characteristic, it describes the fluctuations at length scales larger than the cut-off length *l*. In an equilibrium system, the probability $W_{eq,l}(f)D\varphi(\mathbf{r})$ to find the system in an element $D\varphi$ of the configuration space may be written in the form

$$W_{eq,\lambda}(\phi)D\phi = e^{\frac{F_\lambda - H_l(\phi)}{k_b T}} D\phi, \quad F_l = -k_b T_c \ln \int e^{-\frac{H_l}{k_b T}} D\phi,$$

$$H_l = -F_{reg} - k_b T_c \ln \sum_{(\phi(r))} e^{-\frac{H^{(mic)}}{k_b T}}, \quad F_{reg} = F - F_l. \tag{1}$$

$H^{(mic)}$ is the microscopic "true" Hamiltonian; the effective Hamiltonian of fluctuations $H_l\{\varphi(r);l,T\}$ depends on the cut-off length *l* and on temperature T and other parameters of the thermodynamic state, and *F* is the free energy. A conventional general approach to near-critical fluctuation dynamics is that of the Langevin equations (see ref. [2-3,5]). The general form of these equations is

$$\frac{\partial \varphi_i}{\partial t} = \Gamma_{ik} \{-\frac{\delta H_l}{\delta \varphi_k} + \theta_k\}, \tag{2}$$

$\Gamma_{ik}$ being the kinetic coefficient operator. Extraneous Gaussian random fields $\theta_k(r,t)$ are normalized by the fluctuation-dissipation theorem. For the liquid-gas critical point, equations (2) are taken in the form (model H in the classification of Halperin and Hohenberg[5])

$$\frac{\partial j}{\partial t} = \Im\{\eta_0 \nabla^2 \frac{\delta H_{l_0}}{\delta j} + g(\nabla \varphi) \frac{\delta H_{l_0}}{\delta \varphi} + \zeta_j(r,t)\} = 0,$$

$$\partial \varphi / \partial t = \lambda_{\varphi 0} \nabla^2 \frac{\delta H_{l_0}}{\delta \varphi} - g(\nabla \varphi) \frac{\delta H_{l_0}}{\delta j} + \theta(r,t), \tag{3}$$

$$H = \int dV \{\frac{c_0}{2} (\nabla \varphi)^2 + \frac{\varphi^2}{2\chi_{\varphi 0}} + u_0 \varphi^4 - h\varphi + \frac{1}{2\rho_e} j^2\}.$$

Here, j(r,t) is the momentum current density, and h is proportional to the deviation of the chemical potential from its critical value. The projection operator $\Im$ selects the transverse part of the vector in brackets. The label 0 in the coefficient $c_0$, the shear viscosity $\eta_0$, thermal conductivity $\lambda_0$ and susceptibility $\chi_0$ relates these quantities to an initial length-scale $l_0$. The mode-coupling constant g is introduced for mathematical convenience, its physical value is g=1. The extraneous random sources $\zeta(r,t)$ and $\theta(r,t)$ are Gaussian random fields yielding the relations:

$$<\zeta_j(r,t)\zeta_l(r',t')> = 2(\nabla_j \nabla_l - \delta_{jl}\nabla^2)k_B T\eta_0 \delta(r-r')]^{(tr)}\delta(t-t');$$
$$<\theta(r,t)\theta(r',t')> = -2\nabla^2 k_B T\lambda_{\varphi 0}\delta(r-r')\delta(t-t'); \qquad , \qquad (4)$$
$$<\zeta_j(r,t)> = <\theta(r,t)> = <\zeta_j(r,t)\theta(r,t)> = 0.$$

$k_B$ is the Boltzmann constant; the quantities $\eta_0$ and $\lambda_0$ are functions of the temperature T. The coefficients in (1) and (2) are chosen to satisfy the fluctuation-dissipation theorem [2, 5].

The existence of a finite cut-off length makes the configurations of hydrodynamic fields smooth functions of space coordinates. It follows from general considerations (see Landau and Lifshitz [13]) that thermal fluctuations (and extraneous fields) may be regarded as small: this justifies the approximation of the effective Hamiltonian density in (3) by a polynomial in fields $\varphi$ and their gradients $\nabla\varphi$. The coefficients in these polynomials depend on the cut-off length, they may be calculated by taking averages over the excluded small scale degrees of freedom [see, e.g. 1-3,5]. In the same approximation, the linear operator $\Gamma_{ik}$ of kinetic coefficients may be written as a polynomial in the gradient operator $\nabla$ with only the low order terms usually considered. The polynomial form of the effective Hamiltonian and the kinetic coefficients operator is usually referred to as Landau expansion. The set **L** of coefficients in this expansion represents statistical averages over shorter scales fluctuations. **L** depends on thermodynamic parameters (temperature, pressure) and on the cut-off length $l_0$.

The hydrodynamic description assumes local equilibrium. The local equilibrium picture is not restricted to the global thermodynamic equilibrium state, it is based on the observation that the magnitudes of fluctuations decrease and relaxation times increase with increasing length scale. Consider an initial cut-off scale $l_0$, with relaxation time at this scale $\tau(l_0)$. At times $t \gg \tau(l_0)$, the coefficients $L$ become equilibrium functions of few local thermodynamic parameters, and the hydrodynamic modes yield Langevin equations with the Landau expansion coefficients determined by local thermodynamic parameters. The nonequilibrium state at large length scales is described by these Langevin equations. The temperature T is now treated as one of local thermodynamic parameters defined by small length scale characteristics and the particle kinetic energy playing the role of a heat bath for the larger scale nonequilibrium degrees of freedom.

In a system with temperature gradients, local temperature determines local values of $L(\mathbf{r})$. This, in turn, leads to a new steady state fluctuation statistics at larger scales. A rapid uniform change in temperature results in uniform changes $\delta L$ in the Landau coefficients $L$ at times $\sim \tau(l_0)$ while the larger scale fluctuations have larger relaxation times. Then, the Langevin equations correspond to the new state while the old state of large scale components of the order parameter field give initial conditions for a transient state. One may formally consider all Landau coefficients $L$ (or their deviations from the critical point values) as new independent parameters of state, and describe perturbations leading to nonequilibrium as a perturbations $\delta L$ of these parameters. A small change in most ("irrelevant") coefficients results in small changes in large length scale fluctuations. It is known from the RG that small changes in "relevant" coefficients, $h$ conjugated to the order parameter $\varphi$, and $1/\chi_{\varphi,0}$ conjugated to the energy density, generate changes increasing with increasing length scale. Then, one may neglect changes in "irrelevant" coefficients, by setting them at their critical point values. This allows one to simplify the models for non-equilibrium and non-uniform near-critical states. In following sections, we review the effects in specially selected steady and transient near-critical states that may be used as test situations.

## 4. Near-critical nonequilibrium liquids: mode coupling effect

We consider, following ref. 9, a steady state with a small constant temperature gradient $\nabla T$. For $|\nabla T|<|T-T_c|/L$, $L$ being the size of the system, the only "relevant" term in Langevin equations that "perturbs" the uniform equilibrium model is the mode coupling term $g(j\nabla<\varphi>)$. This term is a spatially uniform but unisotropic perturbation of the critical point regime.

At the very large length scales $R>>r_c$, the fluctuations yield the linearized renormalized Langevin equations

$$\frac{\partial j_i^{(tr)}}{\partial t} = \eta \frac{1}{\rho_e} \nabla^2 j_i^{(tr)} - gc(\alpha_j \nabla_i \nabla_j - \alpha_i \nabla^2)\varphi + \zeta_i;$$

$$\partial \varphi/\partial t = -g \frac{1}{\rho_e} j_i^{(tr)} \alpha_i + \frac{\lambda_\varphi}{\chi_\varphi} \nabla^2 \varphi(r,t) - \lambda_\varphi c \nabla^2 \nabla^2 \varphi + \theta(r,t); \quad\quad (5)$$

$$\alpha_i = \nabla_i <\varphi>; \nabla_i j_i^{(tr)} = 0.$$

In these equations, a cut-off length $l> r_c$ is assumed. The quantities $c(\varepsilon), \eta(\varepsilon), \lambda_\varphi(\varepsilon), \chi_\varphi(\varepsilon)$ are now singular macroscopic characteristics, with general behavior $A(\varepsilon)=A_0\varepsilon^x$ where A denotes the quantity and x is the corresponding critical exponent. Note that $\chi_\varphi(\varepsilon)\sim(\partial\varphi/\partial\varepsilon)_p \sim c_p \sim \varepsilon^{-\gamma}$. The correlation radius is $r_c(\varepsilon)=(c\chi_\varphi)^{1/2}$. We use the non-renormalized order parameter $\varphi$; in the renormalization group, the renormalization of $\varphi$ is used to set $c(\varepsilon)=1$.

For small gradients $\nabla\varepsilon$, $A(\varepsilon(r))=A+(\partial A/\partial\varepsilon)_p(r\nabla\varepsilon)$, where A is the system average. Consider now $\nabla\varepsilon$ as an expansion parameter. The non-dimensional expansion parameter, however, depends on the length-scale r and has the form $(r\nabla A)/A = r\nabla \ln A = xr\nabla \ln \varepsilon$. This parameter becomes of the order of unity at the length-scale $r_{nu} = A/\nabla A$. At this length-scale the variation in A is $\delta A \sim r_{nu}\nabla A = A$, and the non-uniformity of singular properties and of the reduced temperature $\varepsilon$ become large while the non-uniformity of the temperature T is small.

In light scattering experiments (see, e.g. [17]), the dynamical structure factor $S(k,\omega) = <|\varphi(k,\omega)|^2>$ is studied. The Fourier amplitude $\varphi(k,\omega)$ of the order parameter is defined by

$$\varphi(r,t) = \int \varphi_{k,\omega} e^{-i(kr+\omega t)} \frac{d^3k d\omega}{(2\pi)^4}. \tag{6}$$

Fourier amplitudes for other fluctuating fields are defined by similar formulas. From (5) and (4), one obtains the equations for the Fourier-amplitudes

$$(i\omega + \eta k^2)\frac{1}{\rho_e} j^{(tr)}{}_{i;k,\omega} + gck^2 \Delta_{ij}\alpha_j \varphi_{k,\omega} = \zeta_{i;k,\omega};$$

$$(i\omega + D_T k^2)\varphi_{k,\omega} - g\frac{1}{\rho_e} j_{j;k,\omega}\alpha_j = \theta_{k,\omega};$$

$$<\theta_{k,\omega}\theta_{k',\omega'}> = 2(2\pi)^4 k_B T_c \lambda_\varphi k^2 \delta(k+k')\delta(\omega+\omega'). \tag{7}$$

$$<\zeta_{i;k,\omega}\zeta_{j;k',\omega'}> = 2(2\pi)^4 k_B T_c \eta k^2 \Delta_{ij} \delta(k+k')\delta(\omega+\omega');$$

$$k_i j^{(tr)}{}_{i;k\omega} = 0; \Delta_{ij} = \delta_{ij} - \frac{k_i k_j}{k^2}$$

In the frequency range $\omega \sim \omega_\varphi = D_\varphi k^2 << \eta k^2$ of the Rayleigh scattering, one obtains from (7)

$$[i\omega + D_\varphi k^2 + \frac{g^2 \alpha^2 c}{\eta}(1 - \cos^2 \vartheta)]\varphi_{k,\omega} = g\frac{1}{\eta k^2}\alpha_j \zeta_i + \theta_{k,\omega}; \tag{8}$$

$\vartheta$ is the angle between $\alpha$ and k: $(k\alpha) = |k||\alpha|\cos\vartheta$. The structure factor $S_{k,\omega}$ is

$$S_{k,\omega} = <|\varphi_{k,\omega}|^2> =$$

$$= 2(2\pi)^4 \lambda_\varphi k_B T \frac{k^2}{|i\omega + D_T[k^2 + k_{cut}^2(1-\cos^2\vartheta)]|^2}[1 + \frac{k_{mc}^4}{k^4}(1-\cos^2\theta)];$$

$$k_{mc} = \left(\frac{g^2\alpha^2}{\lambda_\varphi \eta}\right)^{\frac{1}{4}} = \left(\frac{g^2(\frac{\partial\varphi}{\partial\varepsilon})_p^2}{\lambda_\varphi \eta}\right)^{\frac{1}{4}} (\nabla\varepsilon)^{\frac{1}{2}}; \quad (9)$$

$$k_{cut} = \left(\frac{g^2\alpha^2 c}{D_T \eta}\right)^{\frac{1}{2}} = g(\frac{\partial\varphi}{\partial\varepsilon})_p \left(\frac{c}{D_T \eta}\right)^{\frac{1}{2}} \nabla\varepsilon.$$

In (9), there are two characteristic wave-vectors, the mode coupling vector $k_{mc} \sim |\nabla\varepsilon|^{1/2}$ and the cut-off vector $k_{cut} \sim |\nabla\varepsilon|$. For small $|\nabla\varepsilon|$, $k_{cut} << k_{mc}$. In the range $1/r_c >> k >> k_{cut}$ one may write

$$S_{k,\omega} = 2D_T k_B T \frac{k^2}{|i\omega + D_T k^2|^2}\left(1 + \frac{k_{mc}^4}{k^4}(1-\cos^2\theta)\right). \quad (10)$$

For liquids far from the critical point, a similar formula was obtained in ref. [14-16], and experimentally tested in ref. [17].

The mode coupling effect becomes dominant for $k < k_{mc}$. The mode-coupling radius $r_{mc} = 1/k_{mc}$ is

$$r_{mc} = \left(\frac{1}{\nabla\varepsilon}\right)^{\frac{1}{2}} \left(\frac{\lambda_\varphi \eta}{(\frac{\partial\varphi}{\partial\varepsilon})_p^2}\right)^{1/4}. \quad (11)$$

The scaling behavior of $r_{mc}$ near the critical point follows from the Einstein-Kawasaki formula [18] $D_T = \lambda_\varphi/\chi_\varphi \sim 1/(\eta r_c)$ and the relation $(\partial\varphi/\partial\varepsilon) \sim c_p \sim \varepsilon^{-\gamma}$, where $c_p$ is the isobaric heat capacity:

$$r_{mc} = r_m \varepsilon^{\frac{1}{4}(\gamma+\nu)} |r_0 \nabla\varepsilon|^{-\frac{1}{2}} \quad (12)$$

$r_m$ is proportional to interparticle distance $r_0$. The critical exponent $(\gamma+\nu)/4 > 0$ is a combination of static critical exponents only; the dynamic exponent disappeared due to the Einstein-Kawasaki relation. The singularity in $\lambda_\varphi(\varepsilon)$ is due to the mode coupling at length-scales in the critical range of scales, so the dynamical exponent is cancelled by a self-consistent mechanism.

The equal-time correlation function $<\varphi(r,t)\varphi(r',t)>$ is related to the structure factor $S_{k,\omega}$:

$$<\varphi(r,t)\varphi(r',t)> = \int S(k,\omega) e^{-ik(r-r')} \frac{d^3k \, d\omega}{(2\pi)^4} \qquad (13)$$

The integral in (13) with the structure factor (9) has a linear divergence at small k:

$$K_d \sim k_B T_c k_{mc}^4 \int_k \frac{dq}{q^2} \sim k_B T_c \varepsilon^{-(\gamma+\nu)} \frac{|\nabla \varepsilon|^2}{k} \quad . \qquad (14)$$

The cut-off vector provided by the formula (8) is $k_{cut} \sim \varepsilon^{-(\gamma+3\nu)/2} |\nabla\varepsilon|$. The applicability of the uniform approximation (5) assumes $k \gg k_{nu} = \nabla\varepsilon/\varepsilon$, but close to the critical point $k_{nu}/k_{cut} \sim \varepsilon^{(\gamma+3\nu-2)/2} \ll 1$. The cut-off provided by $k_{cut}$ gives $K_d \sim \varepsilon^{-(\gamma-\nu)/2}|\nabla\varepsilon|$, so that the contribution vanishes in the limit $\nabla\varepsilon \to 0$.

The mode-coupling length $r_{mc}$ decreases and the correlation radius $r_c$ increases when $\varepsilon$ decreases at a constant $|\nabla\varepsilon| \neq 0$. These two characteristic lengths coincide at $\varepsilon=\varepsilon_{cn}$ yielding the condition

$$r_0 \nabla \varepsilon_{cn} = (\frac{r_m}{r_0})^2 \varepsilon_{cn}^{\frac{\gamma+5\nu}{2}} \quad . \qquad (15)$$

The variation $\delta\varepsilon$ of the reduced temperature at the length scale $r_c=r_{mc}$ is $\delta\varepsilon/\varepsilon=(r_c\nabla\varepsilon)/\varepsilon=(r_m^2/r_0)\varepsilon_{cn}^{(\gamma+3\nu)/2-1}$. For the values $\gamma=1.239$, $\nu=0.630$ of the critical exponents [19] one has $\delta\varepsilon/\varepsilon \sim \varepsilon_{cn}^{0.565} \ll 1$. Then, at the length scale $r_c=r_{mc}$ the non-uniformity is small, and the approximation (5) applicable.

In the plane (e-$\nabla\varepsilon$), the condition $\varepsilon=\varepsilon_{cn}(\nabla\varepsilon)$ determines the crossover from weak gradient to strong gradient regime. We assume that in the strong gradient regime $\varepsilon<<\varepsilon_{cn}(\nabla\varepsilon)$ the mode coupling effect disrupts the critical fluctuations at length-scales $r\geq r_{mc}(\nabla\varepsilon)$. Then, the $\nabla\varepsilon$-dependent upper limit of the critical range of scales is the new correlation radius $r_c(\nabla\varepsilon)=r_{mc}(\nabla\varepsilon)$. Note that the singularities of thermodynamic and kinetic characteristics appear in the theory [1-3,5] as integrals of correlation functions over length scales. At the critical point, these integrals diverge at large scales. In a near-critical state, the large but finite correlation radius $r_c$ provides for a cut-off, and the singular characteristics may be written as power functions of $r_c$. We assume that this is applicable to the new correlation radius $r_c(\nabla\varepsilon)$ and to singular characteristics entering the formula (9) for the mode coupling radius $r_{mc}$. Then, the condition $r_c(\nabla\varepsilon)=r_{mc}(\nabla\varepsilon)$ gives for $r_c(\nabla\varepsilon)$ at constant $\varepsilon<\varepsilon_{cn}$

$$r_c(\nabla\varepsilon) = r_0 \left(\frac{r_0}{r_m}\right)^{\frac{4\nu}{\gamma+5\nu}} (r_0\nabla\varepsilon)^{-\frac{2\nu}{\gamma+5\nu}} \sim (\nabla\varepsilon)^{-0.287}. \qquad (16)$$

The new scaling lows for singular characteristics follow from their dependence on $r_c$ and formulas (13)-(15). This may be used to test the theoretical prediction (16) in thermodynamic experiments. A test of the basic assumptions of this theory is possible by a comparative study of the non-equilibrium mode coupling effects caused by temperature gradient and of effects caused by an equilibrium order parameter gradient $\nabla<\varphi>\neq 0$ imposed by an external field h(r), for example by the earth gravity.

## *5. Transient near-critical states*

In this section, transient near-critical states created by a rapid, spatially uniform temperature or other parameter change are discussed following ref. 11,12. The time $t_{ch}$ of the rapid change is supposed to be large when compared to the relaxation time $\tau(l_0)$, but $t_{ch}$ is negligibly small when compared to $\tau(R)$, R being the large scale of observation. At times $t > t_{ch} > \tau(l)$, fluctuations at the initial and smaller scales have reached the new equilibrium state, while the larger scales only began the relaxation. The coefficients L, and Langevin equations for larger scale fluctuations are determined by the initial and smaller scales fluctuations. The kinetics at large length scales R having relaxation times $\tau(R) >> t$ may be described as relaxation of the initial conditions determined by the state before temperature change but governed by equations corresponding to the final state. This argument is based on local equilibrium ideas. Note that the most of degrees of freedom, and the main contribution to the energy and entropy of the system, belong to the atomic length scale $r_0$. The atomic scale plays the role of a "thermal bath" for the rest of the degrees of freedom. At times $t >> \tau(r_0)$ (for room temperatures and atomic scale $r_0 \sim 10^{-7}$ cm, $\tau(r_0) \sim 10^{-12}$ sec) one characterizes this "thermal bath" by a time-dependent temperature T(t). This is the temperature measured by devices thermally coupled to the liquid.

One has to study separately a step towards the critical point ($r_{c,fin} >> r_{c,in}$) in which a new range of strongly interacting fluctuations appear, and a step away from the critical point ($r_{c,fin} << r_{c,in}$) where a large range of critical fluctuations disappears. Following the dynamic RG method [2,5], we introduce a variable scale $\lambda >> r_0$. At times $t \sim \theta(\lambda)$, the fluctuations on length scales $R < \lambda$ are in the new equilibrium state.

Consider first a temperature step away from the critical point [11], so that $r_{c,fin} << r_{c,in}$. Most important are changes in the fluctuation picture in the range of scales $r_{c,fin} < r < r_{c,in}$; we chose the length $\lambda$ in this range. At $t > \theta(l) > \theta(r_{fin})$ the scales $r < \lambda$ have approached the new equilibrium characterized by the new correlation radius $r_{fin}$, and the RG calculation at $t>\theta(r_{fin})$ gives $H_\lambda$ as a free field Hamiltonian

$$H_\lambda = \sum_{k<k_\lambda} H_k \, , \quad H_k = \frac{1}{2\chi_{fin}}|\phi_k|^2 \, . \tag{17}$$

The reduction of the problem to that of a free field allows one to exactly solve the problem of long-range fluctuations without further assumptions. For the time-dependent average $M_{k(t)}=<|\phi_k(t)|^2>$ one obtains

$$M_k(t) = (M_{k,in} - M_{k,fin})e^{-\frac{2t}{\tau_k}} + M_{k,fin}, \quad M_{k,in} = <|\phi_k|^2>_{in}, \quad M_{k,fin} = 2k_b T_c \chi_{fin}; \quad \tau_k = \frac{1}{D_{fin} k^2}. \tag{18}$$

The diffusion coefficient $D_{fin}$ is scale-independent. At times $t<t_k$, $M_k(t)>>M_{k,fin}$, and the probability distribution at length scale $R\sim 1/k$ retains the initial non-Gaussian shape. At $t> t_k$ the fluctuations on the scale $\lambda\sim 1/k$ approach equilibrium and become Gaussian; for more details see ref. 11.

The rapid increase of the temperature requires more energy than a slow heating. The effective Hamiltonian $H_\lambda(\phi)$ is by definition the free energy of the system in a state with given $\phi_k$, $k<1/\lambda$. The entropy in this state may be written as $S_\lambda=-\partial H/\partial T=(1/T_c)<H_k>(\partial \ln\chi/\partial\varepsilon)$. The contribution of a single harmonic $\phi_k$ to the entropy is $s_k=(1/T_c)<H_k>(\partial \ln \chi/\partial\varepsilon)$. The single harmonic internal energy $U_k$ is (according to the thermodynamic formula $U=F+TS$ ) $U_k=<H_k>+T_c s_k$ . Let the initial state be at the critical point $\varepsilon_{in}=0$, and $<|\phi_k|^2>_{in}\sim k^{-2+\eta}$ . In the final equilibrium state one finds $<|\phi_k|^2>_{fin} \sim \chi(\varepsilon_{fin})$, $k<<1/r_{c,fin}$ . Summing up the contributions from all large-scale harmonics, one obtains the excess energy $U_{stor}$ stored at $t\sim\theta(r_{c,fin})$:

$$U_{stor}(0) = \frac{V}{2}\frac{\partial \ln \chi_{fin}}{\partial \varepsilon}\frac{1}{\chi_{fin}}\int_0^{1/\lambda}<|\phi_k|>_{in}^2 \frac{d^3k}{(2\pi)^3} \approx Nk_bT_c\varepsilon_{fin}^{3\nu-1}. \quad (19)$$

Here $\lambda \sim r_{c,fin}$, and $N \sim V/r_0^3$ is the number of molecules in the system. At $t > \theta(r_1)$ the stored energy is gradually converted into heat. The released heat $Q(t) = U_{stor}(0) - U_{stor}(t)$ as a function of time is

$$Q(t) = \frac{V}{2}\frac{\partial \ln \chi}{\partial \varepsilon_{fin}}\frac{1}{\chi_{fin}}\int_0^{k_\lambda}[<|\phi_k(0)|^2>_- <|\phi_k(t)|^2>]\frac{d^3k}{(2\pi)^3} \propto U_{stor}(0)[1-(\frac{\tau(r_{c,fin})}{t})^{\frac{1+\eta}{2}}]. \quad (20)$$

So far we have considered the isothermal relaxation at $\varepsilon(t) = \varepsilon_{fin}$. In a system adiabatically insulated at $t > \theta(r_{fin})$, the time-dependent temperature yields the relation $dQ = C_v dT = CT_c d\varepsilon(t)$, with $C_v \sim \varepsilon^{3n-2}$ being the heat capacity at constant volume. The approach of the temperature to equilibrium follows now from the law (25) of heat release. For $t > \theta(r_{c,fin})$ one obtains

$$\varepsilon(t) - \varepsilon_{fin} \_ \varepsilon_{fin}[\tau(r_{c,fin})/t]^{\zeta_+}, \quad \zeta_+ = (1+\eta)/2. \quad (21)$$

Let us now consider a step toward the critical point [12]: $\varepsilon_{in} \gg \varepsilon_{fin}$. In contrast to the above case, the harmonics $\varphi_k$ in the range $r_{c,in} < 1/k < r_{c,fin}$ strongly correlate in the final state, and the problem cannot be reduced to that of a single harmonic. The suggested scenario is sequential equilibration: at a given time t, the fluctuations on small scales up to a time-dependent equilibration length scale $\lambda_{eq}(t)$ will equilibrate while the harmonics $\varphi_k$ with $k < 1/\lambda_{eq}$ will retain the initial magnitudes. These magnitudes are small compared to the final ones, and may be neglected. The statistics and the kinetics of the fluctuations on scales $r < \lambda_{eq}(t)$ yield then the equilibrium scaling relation: on a scale $\lambda < \lambda_{eq}(t)$ the relaxation time is $\theta(\lambda) \sim \lambda^2/D(\lambda) \sim \lambda^{3+z/n}$, where z is the critical exponent for the viscosity[19]. The time dependence of $\lambda_{eq}$ is controlled by the condition that $\theta(R) < t$ for $r < \lambda_{eq}(t)$, and $\theta(R) > t$ for $r > \lambda_{eq}(t)$. The scale $\lambda_{eq}(t)$ has then the equilibration time $\theta(\lambda_{eq}(t)) \sim t$; this relation gives the scaling law for $\lambda_{eq}(t)$

$$\lambda_{eq}(t) = r_{in}[t/\tau(r_{in})]^p, \quad p = \frac{\nu}{3\nu + z}. \tag{22}$$

$\nu \cong 0.630$ and $z \cong 0.063$ (see ref. 19) give $p \cong 0.32$.

The sequential "cooling" of large scale fluctuations is accompanied by the release of heat $Q(t)$. The heat released on the scale $l_{eq}$ is transferred to smaller scales and through the small scale "heat bath" to the thermostat that maintains the temperature $T(t)$=constant. The scaling formula for the entropy $S(\lambda_{eq})$ of fluctuations on scales $r<\lambda_{eq}$ is $S_c - S(\lambda_{eq}) \sim (\lambda_{eq}/r_0)^{(1-3\nu)/\nu}$. The energy conservation law $dQ=TdS$ results in

$$Q(t) = T_c[S(t) - S(\theta(r_{in}))], \quad Q_{fin} - Q(t) \propto Q_{fin}[\theta(r_{c,in})/t]^{(3\nu-1)/(3\nu+z)}, \quad t > \theta(r_{c,in}), \tag{23}$$

where $Q_{fin} \sim T_c \varepsilon_{in}^{3\nu-1}$ is the total amount of heat released.

Consider now a system that is adiabatically insulated during relaxation. At $t>t_1$, the heat $Q(t)$ transferred from the large to the smaller scales $R<\lambda_{eq}$ will result in an increase in the temperature $T(t)$. Then, a rapid quench results in a temperature minimum. The heat capacity $C_v(t_1)$ of the cooling at a time scale $t_1$ is a fraction $x$, $0<x<1$, of the equilibrium heat capacity $C_v$ because the rapid cooling leaves the large scale fluctuations unchanged. From scaling arguments, one writes $C_v \cong A\varepsilon^{3\nu-2} = A(r_c/r_0)^{(2-3\nu)/\nu}$, $C_v(t_1) \cong xA(\lambda_{eq}(t_1)/r_0)^{(2-3\nu)/\nu}$. The conserved internal energy of the adiabatically insulated system is $U_{fin}=U_{in}-DU$, $DU=C_v(t_1)(T_{in}-T_1) \cong C_v(t_1)T_c(\varepsilon_{in}-\varepsilon_1)$. A quasistatic cooling characterized by the same decrease $DU$ of the internal energy determines the final equilibrium state and the difference $\varepsilon_{in}-\varepsilon_{fin} \sim x(\varepsilon_{in}-\varepsilon_1)$, where $\varepsilon_1=(T_1-T_c)/T_c$ and $T_1$ is the temperature at the insulation time $t=t_1$. To arrive at $0<\varepsilon_{fin}<<\varepsilon_{in}$, the temperature $T_1$ must be below $T_c$ so that $\varepsilon_1<0$, $|\varepsilon_1|\sim\varepsilon_{in}$. The phase separation at temperatures $\varepsilon(t)<0$ is avoided if at any time the critical nucleus has the size $R_c(t)>\lambda_{eq}(t)$. The time-dependent heat capacity of the system is $C_v(t) \sim xA[\lambda_{eq}(t)/r_0]$, with $\lambda_{eq}(t)$ given by formula (22). From the thermodynamic relation $dQ(t)=C_v(t)dT=C_v(t)T_c d\varepsilon(t)$ one obtains the scaling law

$$\varepsilon(t) \propto \varepsilon_1 [\theta(r_{c,in})/t]^{\zeta_-}, \quad \zeta_- = 1/(3\nu + z). \quad (24)$$

The critical exponent $\zeta_-=1/(3n+z)$ differs from $\zeta_+$ given by (27). Using the theoretical values for the critical exponents (see ref. 19), one obtains for the liquid-gas and for the binary mixtures critical points $\zeta_+ =0.517$ and $\zeta_- =0.512$. Both exponents are close to each other and to the mean-field value ½.

**Summary**

A theory, based on scaling assumptions, predicts new critical phenomena and scaling lows for nonequilibrium steady and transient near-critical states. Some of the hypotheses are likely to fail in special conditions. For example, the Landau coefficient may become nonuniform following a uniform change of parameters. This instability is expected when a perturbation pushes the system in the two-phase region of the phase plane. Experimental tests of the theory may prove the applicability of the normal scaling hypothesis to a more general than uniform equilibrium conditions.

**Acknowledgements**

It is a pleasure to acknowledge R.W.Gammon, A. Onuki, M.A.Ratner, and R. Allen Wilkinson for helpful discussions. This study was supported by a NASA grant (GRANT NAG3-1932).